\documentclass[prb, preprint, twocolumn, amsmath, amssymb, reprint, superscriptaddress,]{revtex4-1}

\usepackage{graphicx,verbatim}
\usepackage{dcolumn}
\usepackage{bm}
\usepackage{sidecap}
\usepackage{braket}
\usepackage{color}

\begin{document}

\title{Hierarchical quantum master equation approach to charge transport in molecular junctions with time-dependent molecule-lead coupling strengths}
\author{A.\ Erpenbeck}
\affiliation{
Institut f\"ur Theoretische Physik und Interdisziplin\"ares Zentrum f\"ur Molekulare 
Materialien, \\
Friedrich-Alexander-Universit\"at Erlangen-N\"urnberg,\\ 
Staudtstr.\, 7/B2, D-91058 Erlangen, Germany
}
\author{L.\ G\"otzend\"orfer}
\affiliation{
Institut f\"ur Theoretische Physik und Interdisziplin\"ares Zentrum f\"ur Molekulare 
Materialien, \\
Friedrich-Alexander-Universit\"at Erlangen-N\"urnberg,\\ 
Staudtstr.\, 7/B2, D-91058 Erlangen, Germany
}
\author{C.\ Schinabeck}
\affiliation{
Institut f\"ur Theoretische Physik und Interdisziplin\"ares Zentrum f\"ur Molekulare 
Materialien, \\
Friedrich-Alexander-Universit\"at Erlangen-N\"urnberg,\\ 
Staudtstr.\, 7/B2, D-91058 Erlangen, Germany
}
\affiliation{
Physikalisches Institut, Albert-Ludwigs-Universit\"at Freiburg, \\
Hermann-Herder-Strasse 3, 79104 Freiburg, Germany
}
\author{M.\ Thoss}
\affiliation{
Institut f\"ur Theoretische Physik und Interdisziplin\"ares Zentrum f\"ur Molekulare 
Materialien, \\
Friedrich-Alexander-Universit\"at Erlangen-N\"urnberg,\\ 
Staudtstr.\, 7/B2, D-91058 Erlangen, Germany
}
\affiliation{
Physikalisches Institut, Albert-Ludwigs-Universit\"at Freiburg, \\
Hermann-Herder-Strasse 3, 79104 Freiburg, Germany
}

\date{\today}

\begin{abstract}
	Time-dependent currents in molecular junctions can be caused by structural fluctuations or interaction with external fields. In this publication, we demonstrate how the hierarchical quantum master equation approach can be used to study time-dependent transport in a molecular junction. This reduced density matrix methodology provides a numerically exact solution to the transport problem including time-dependent energy levels, molecule-lead coupling strengths and transitions between electronic states of the molecular bridge. Based on a representative model, the influence of a time-dependent molecule-lead coupling on the electronic current is analyzed in some detail. 
\end{abstract}

\maketitle

\section{Introduction}
% 	Molecular junctions in general
	Investigating quantum transport through nanostructures combines the possibility to study fundamental aspects of non-equilibrium many-body quantum physics at the nanoscale with the perspective for applications in nanoelectronic devices. \cite{Nitzan2001, Joachim2205, Cuniberti, Cuevas_Scheer, Thoss2018}
	Studies of transport in quantum dots or molecular junctions have revealed a variety of interesting transport phenomena, such as rectification, switching, quantum interference, or negative differential resistances.\cite{Gaudioso2000, WuNazinHo04, Blum05, Pop2005, Elbing05, Choi2006, Solomon2008_2, Quek2009, Meded09, Diez2009, Benesch2009, Molen2010, Markussen2010, Heinrich2001, Rainer2011, Haertle11_3, Ballmann2012, Guedon2012, Rainer2013, Erpenbeck2015, Erpenbeck2016}
	Along these lines, time-dependent transport has attracted increasing interest recently as transient phenomena are becoming accessible experimentally.\cite{Selzer2013,Cocker2016}

% 	Experimental significance
	Understanding the dynamical response of a molecular junction is essential for prospective applications, but also provides insight into the transport physics.
	The effect of an external driving field on transport in nanosystems was studied extensively,\cite{Bruder1994, Grafstroem2002_Review, Platero2004_Review, Kohler2005, Franco2008, Cuevas_Scheer, Galperin2012_Review} 
	giving rise to the concept of photon-assisted tunneling and
	phenomena such as 
	coherent destruction of tunneling.\cite{Tien1963, Grossmann1991, Bruder1994, Blick1995, Oosterkamp1997, Pedersen1998, Grafstroem2002_Review, Tikhonov2002, Kohler2005, Kleinekathoefer2006, Viljas2007, Meyer2007, Li2007, Li2008}
	This was further extended to schemes that allow to control the current across a molecular junction by laser pulses and studies of the effect of intra-molecular transitions on the current.\cite{Ovchinnikov2005, Welack2006, Kleinekathoefer2006, Kohler2007, Viljas2007, Li2007, Franco2008, Li_2_2008, Kohler2010, Volkovich2011, Fainberg2011, Park2012, Selzer2013, Ochoa2015}
	Based on that, molecular electronic devices such as rectifiers, switches or pumps were proposed that operate by an external field driving the junction.\cite{Lehmann2002, Lehmann2003, Lehmann2003_282, Tu2006, Thanopulos2007, Fainberg2007, Prociuk2010, Kwapinski2011, Peskin2012, Selzer2013}

% 	Theoretical challenges
	A variety of theoretical approaches are available to describe electron transport across a molecular junction in the presence of external driving.\cite{Jauho1994, Pedersen1998, Platero2004_Review, Kohler2005, Peskin2016, Yang2017_Review}
	Thereby, the effect of a time-dependent field is described by time-dependent parameters of the Hamiltonian. 
	Most theoretical studies have focused on the influence of time-dependent energies of the molecular electronic states, which account for phenomena such as photon-assisted tunneling.\cite{Bruder1994, Sun1998, Welack2006, Kleinekathoefer2006, Li2007}
	Moreover, the effect of time-dependent couplings between electronic states that are either localized on the molecule or the leads was investigated extensively within the realm of field-induced electronic transitions and pump-probe setups.\cite{Kohler2010, Fainberg2011, Volkovich2011, Peskin2012, Selzer2013, Ochoa2015}
	The coupling between the molecule (or a quantum dot) and the leads, which enables transport in the first place, can also vary in time. This effect was studied theoretically, e.g., in the context of electron shuttles,\cite{Gorelik1998, Kaun2005, Jorn2009} charge pumps\cite{Splettstoesser2005, Braun2008, Cavaliere2009, Romeo2010, Croy2012, Croy2012_b, Kwapinski2011} and turnstile devices.\cite{Liu1993, Moldoveanu2007, Potanina2017}
	Time-dependent molecule-lead couplings can also arise in the semi-classical limit as a consequence of current-induced dynamics,\cite{Gorelik1998, Kaun2005, Toroker2007, Jorn2009, Pozner2014, Erpenbeck2018}
	or formally by a unitary transformation of time-dependent energy levels in the leads.\cite{Peskin_Baldea, Peskin2016}

% 	Problem adressed in this paper
	In this work, we discuss the physical differences between various time-dependent parameters based on minimal models, with a special focus on time-dependent molecule-lead coupling strengths.
	To this end, we apply a transport theory that is capable of treating a variety of different time-dependent model parameters on the same footing, namely the numerically exact hierarchical quantum master equation approach (HQME).\cite{Tanimura1989, Tanimura2006, Yan2008, Zheng2009, Zheng2012, Li2012, Zheng2013, Haertle2013a, Haertle2014, Haertle2015, Cheng2015, Dattani2015, Ye2016, Wenderoth2016, Schinabeck2016}

% 	Outline
	The outline of this paper is as follows: After introducing the model system in Sec.\ \ref{sec:model}, we explore possible relations between the different time-dependent model parameters in Sec. \ref{sec:time-dep}.
	Sec.\ \ref{sec:HQME} introduces the HQME approach to electron transport and demonstrates how time-dependent molecule-lead couplings can be included in this numerically exact framework. 
	In Sec.\ \ref{sec:results}, we study a representative model and analyze transport processes that depend on the molecule-lead coupling strength. Sec.\ \ref{sec:conclusion} provides a conclusion.

\section{Theoretical methodology}
	\subsection{Model Hamiltonian}\label{sec:model}
		We consider a single-molecule junction, that is a molecule chemically bound to two macroscopic leads, described by the Hamiltonian
		\begin{eqnarray}
			H(t)	&=&	H_\text{M}(t) + H_\text{L}(t) + H_\text{R}(t) + H_\text{ML}(t) + H_\text{MR}(t). \nonumber \\ \label{eq:Hamiltonian}
		\end{eqnarray}
		In principle, any part of the overall system can be time-dependent as indicated by the dependence on $t$.
		In this work, we describe the molecule by a minimal model system consisting of a single electronic state with time-dependent energy $\epsilon_0(t)$, 
		\begin{eqnarray}
			H_{\text{M}}(t)   &=& \epsilon_0(t) d^\dagger d, 
		\end{eqnarray}
		where $d^{\dagger}$/$d$ denote the electronic creation and annihilation operators, respectively. A generalization to multiple electronic states is straightforward.
		The left and right leads are described by a continuum of noninteracting electronic states with time-dependent energies $\epsilon_k(t)$,
		\begin{eqnarray}
			H_\text{L/R}(t)	&=&	\sum_{k\in\text{L/R}} \epsilon_k(t) c_k^\dagger c_k ,
		\end{eqnarray}
		with $c_k$/$c_k^{\dagger}$ being the creation and annihilation operators of the electronic lead states.
		In order to allow for transport, the molecule interacts with the fermionic environment given by the leads, as described by the Hamiltonian
		\begin{eqnarray}
			H_\text{ML/R}(t)	&=&	\sum_{k\in\text{L/R}} \left( V_k(t) c_k^\dagger d  + \text{h.c.} \right), \label{eq:Hamiltonian_coupling}
		\end{eqnarray}
		giving rise to the time-dependent spectral density $\Gamma_\text{L/R}(t, t', \epsilon) = 2\pi\sum_{k\in\text{L/R}} V_k(t) V_k^*(t') \delta(\epsilon-\epsilon_k)$. In the remainder of the paper, we will exclusively work in the wide-band limit, that is assuming that $\Gamma_\text{L/R}$ is energy-independent.

	 \subsection{Mapping between time-dependent model parameters}\label{sec:time-dep}
		Time-dependent Hamiltonians of the form of Eq.\ (\ref{eq:Hamiltonian}) were used extensively for modeling transport through a molecular junction in the presence of external influences, such as electric fields. Most of the theoretical studies concentrated on the time-dependencies entering the molecular Hamiltonian $H_\text{M}(t)$ or the Hamiltonian describing the leads $H_\text{L/R}(t)$. A common approach for understanding time-dependent transport is the concept of photon-assisted tunneling,\cite{Tien1963, Blick1995, Oosterkamp1997, Pedersen1998, Grafstroem2002_Review, Viljas2007, Meyer2007} 
		which is based on considering time-dependent energies and gives rise to phenomena such as the coherent destruction of tunneling.\cite{Grossmann1991, Kohler2005, Li2007, Li2008} 
		Transitions between molecular electronic states, which can induce important transport channels for a molecule influenced by an electric field, are described by time-dependent couplings between states in $H_{\text{M}}(t)$.\cite{Kohler2010, Volkovich2011, Peskin2012, Selzer2013, Ochoa2015} Along these lines, model systems acting like switches, rectifiers or pumps were identified.\cite{Lehmann2002, Lehmann2003, Lehmann2003_282, Tu2006, Thanopulos2007, Prociuk2010, Kwapinski2011, Peskin2012, Selzer2013}
		
		In contrast to that, time-dependent molecule-lead coupling strengths have received less attention outside the realm of electron shuttles, charge pumps and turnstile devices.\cite{Gorelik1998, Kaun2005, Jorn2009, Splettstoesser2005, Braun2008, Cavaliere2009, Romeo2010, Croy2012, Croy2012_b, Liu1993, Moldoveanu2007, Potanina2017} 
		In this section, we discuss the physics inherent to time-dependent molecule-lead coupling strengths.

		Generally, time-dependent energies, intra-molecular couplings and molecule-lead couplings are not independent. However, there is no one-to-one correspondence as we will demonstrate in the following.
		Using an unitary transformation it is possible to map time-dependent energies,\cite{Peskin_Baldea, Peskin2016} and under certain conditions also time-dependent intra-molecular couplings,\cite{Peskin2012, White2013} to a time-dependent phase factor multiplying the molecule-lead coupling strength.
		Without loss of generality, we consider the scenario of time-dependent energies in the left lead, 
		\begin{eqnarray}
			H(t) &=& 	\epsilon_0 d^\dagger d + \sum_{r\in\text{R}} \epsilon_r c_r^\dagger c_r + \sum_{l\in\text{L}} \epsilon_l(t) c_l^\dagger c_l \nonumber \\ &&
				  +	\left[\sum_{r\in\text{R}} V_r c_r^\dagger d +\sum_{l\in\text{L}} V_l c_l^\dagger d + \text{h.c.} \right].
		\end{eqnarray}
		Peskin\cite{Peskin2016} showed that using an unitary transformation acting on an arbitrary operator $O$,
		\begin{eqnarray}
		\tilde O &=& e^{\frac{i}{\hbar} S(t)} O e^{-\frac{i}{\hbar}S^\dagger(t)},
		\end{eqnarray}
		with 
		\begin{eqnarray}
		S(t)&=& \sum_{l\in\text{L}} c_l^\dagger c_l \int_0^t \epsilon_l(\tau)-\epsilon_l(0) d\tau,
		\end{eqnarray}
		the time-evolution of the transformed density matrix $\tilde\rho(t)$ is governed by the Hamiltonian
		\begin{eqnarray}
			\tilde H(t) &=& 	\epsilon_0 d^\dagger d + \sum_{r\in\text{R}} \epsilon_r c_r^\dagger c_r + \sum_{l\in\text{L}} \epsilon_l(0) c_l^\dagger c_l \\ &&
					+	\left[\sum_{r\in\text{R}} V_r c_r^\dagger d +\sum_{l\in\text{L}} V_l e^{\frac{i}{\hbar} \int_0^t \epsilon_l(\tau)-\epsilon_l(0) d\tau} c_l^\dagger d + \text{h.c.} \right]. \nonumber
		\end{eqnarray}
		Accordingly, a time-dependent energy, which is necessarily a real function of time, can be considered as a time-dependent phase factor multiplying the molecule-lead coupling strength.
		
		Inversely, it is not always possible to transform a time-dependent molecule-lead coupling $V_k(t)$ into a time-dependent energy.
		A coupling strength $V_k(t)$ with a time-varying absolute value does not map onto time-dependent energy-levels as its time-dependency can not be expressed by a single time-dependent phase factor alone.
		Consider, for example, the same transformation as above, using 
		\begin{eqnarray}
		S(t)&=& -i\hbar\sum_{l\in\text{L}} c_l^\dagger c_l \ln(V_l(t)/V_l(0)). 
		\end{eqnarray}
		Although the time-dependency of $V_l(t)$ can be mapped mathematically onto the lead energies, the transformation is not unitary and introduces complex energies with a time-dependent imaginary part, which contradicts a Hamiltonian formulation. 

		To emphasize the physical difference between time-dependent energies and coupling strengths, consider the influence of a harmonic modulation with frequency $\omega$.
		A harmonically oscillating energy-level $\epsilon_0(t) \propto \cos(\omega t)$ gives rise to a multitude of transport channels, which are displaced by multiples of the energy of the driving frequency $n\hbar\omega$, $n\in \mathbb{Z}$.
		This was demonstrated, for example, by Peskin,\cite{Peskin2016} but also realized by other authors.\cite{Cuevas_Scheer, Tien1963, Tucker1985, Kohler2005, Peskin_Baldea}  The different transport channels correspond to the emission or absorption of an integer number of photons with the respective probability given by Bessel functions.\cite{Cuevas_Scheer, Peskin2016}

		If we consider, on the other hand, an oscillating molecule-lead coupling strength,
		\begin{eqnarray}
			H(t) &=& 	\epsilon_0 d^\dagger d + \sum_{r\in\text{R}} \epsilon_r c_r^\dagger c_r + \sum_{l\in\text{L}} \epsilon_l c_l^\dagger c_l \nonumber \\ &&
					+	\left[\sum_{r\in\text{R}} V_r c_r^\dagger d +\sum_{l\in\text{L}} V_l e^{\pm i\omega t} c_l^\dagger d + \text{h.c.} \right], \label{eq:H_perioduc_V_k}
		\end{eqnarray} 
		then the influence of the driving field can be mapped onto a time-independent Hamiltonian
		\begin{eqnarray}
			\tilde H &=& 	\epsilon_0 d^\dagger d + \sum_{r\in\text{R}} \epsilon_r c_r^\dagger c_r + \sum_{l\in\text{L}} \left[\epsilon_l\pm\hbar\omega\right] c_l^\dagger c_l \nonumber \\ &&
				  +	\left[\sum_{r\in\text{R}} V_r c_r^\dagger d +\sum_{l\in\text{L}} V_l c_l^\dagger d + \text{h.c.} \right].
		\end{eqnarray}
		Remarkably, this Hamilton suggests the existence of a single photon process only. 
		Physically, the time-dependent molecule-lead interaction in Eq.\ (\ref{eq:H_perioduc_V_k}) describes a Rabi-type coupling between the electronic state of the molecule and the electronic states in the lead, which results in an oscillation of electrons between the molecule and the lead. 
		Consequently, the time-dependency of the molecule-lead coupling is most influential for electronic states in the leads fulfilling the resonance condition $\epsilon_0-\epsilon_l=\pm\hbar\omega$.
		The oscillation of charge between the molecular electronic state and these lead states meeting the resonance condition is most pronounced and occurs with a Rabi frequency which is smaller than for cases at which the resonance condition is not fulfilled. 
		In this case, the charge resides during a Rabi cycle long enough on the molecule, such that it can be influenced by the presence of the other lead and contribute to transport, giving rise to a single resonance process only.
		
		In conclusion, time-dependent molecule-lead coupling strengths represent a fundamental time-dependency entering the Hamiltonian. As it is not always possible to express their influence by other time-dependent parameters, it is necessary to derive a transport theory that is capable to describe any time-dependency, including time-dependent molecule-lead couplings, on the same footing.
		To this end, we derive in the next section a hierarchical quantum master equation (HQME) approach that can describe the influence of time-dependent molecule-lead coupling strengths.

	\subsection{Transport theory}\label{sec:HQME}
		The HQME method, also known as hierarchical equations of motion (HEOM), is a numerically exact density matrix based method for the description of dynamics and transport in open quantum systems. It was originally developed by Tanimura and Kubo to describe relaxation dynamics in molecular quantum systems,\cite{Tanimura1989, Tanimura2006} and was extended to the simulation of non-equilibrium electron transport by Yan \textit{et al.}\cite{Yan2008, Zheng2009, Zheng2012, Li2012, Zheng2013, Cheng2015, Ye2016} and H\"artle \textit{et al.}.\cite{Haertle2013a, Haertle2014, Haertle2015, Wenderoth2016, Schinabeck2016}
		The HQME method generalizes perturbative master equation approaches by including higher-order contributions as well as non-Markovian memory and allows for the systematic convergence of the results. Being a time-local generalized master equation approach, it is particularly well suited to describe time-dependent effects in transport.
		A detailed derivation of the HQME method in the context of transport is given in Refs.\ \onlinecite{Yan2008, Zheng2012, Haertle2013a}.

		The HQME theory describes the dynamics of open quantum systems, separating the overall problem into a system and a bath. In the context of molecular junctions, the molecule is considered as the system, whereas the leads represent the bath.
		The HQME approach provides an equation of motion for the $n$th-tier auxiliary density operators $\rho_{a_1 \dots a_n}^{(n)}(t)$, 
		\begin{eqnarray}
			\frac{\partial}{\partial t} \rho_{a_1 \dots a_n}^{(n)}(t) 	&=& 
			-\frac{i}{\hbar} [H_\text{M}(t), \rho_{a_1 \dots a_n}^{(n)}(t)] \nonumber \\&&
			-\left( \sum_{j=1}^n \gamma_{a_j}(t) \right) \rho_{a_1 \dots a_n}^{(n)}(t) \nonumber \\&&
			-i \sum_{j=1}^n (-1)^{n-j} \mathcal{C}_{a_j}(t) \rho_{a_1 \dots a_{j-1} a_{j+1} \dots a_n}^{(n-1)}(t) \nonumber \\&&
			-\frac{i}{\hbar^2} \sum_{a_{n+1}} A_{K_{n+1}}^{\overline{\sigma_{{n+1}}}}(t) \rho_{a_1 \dots a_n a_{n+1}}^{(n+1)}(t) .
			\label{eq:EQM_nth_tier}
		\end{eqnarray}
		The zeroth-tier auxiliary density operator represents the reduced density matrix of the system, $\rho(t)$, the higher tier auxiliary density operators describe the influence of the environment on the dynamics of the system.
		The $n$th-tier auxiliary density matrices have $n$ compound indices $a_j = (K_j, p_j, \sigma_j)$, including a lead index $K_j\in \lbrace \text{L}, \text{R} \rbrace$ and an index corresponding to the molecular creation/annihilation operator $\sigma_j \in \lbrace +,- \rbrace$. 
		For a molecule-lead coupling of the form of Eq.\ (\ref{eq:Hamiltonian_coupling}), all information about the influence of the leads on the molecule is encoded in the two-time correlation function
		\begin{eqnarray}
			C_K^\pm(t, t') &=& \sum_{k \in K} V_k(t) V_k^*(t') \braket{ F^\pm_{Kk}(t) F^\mp_{Kk}(t') },
		\end{eqnarray}
		with the operators 
		\begin{eqnarray}
			F^\pm_{Kk}(t) &=& e^{\frac{i}{\hbar} \int_0^t H_K(\tau) d\tau } c_k^\pm e^{-\frac{i}{\hbar} \int_0^t H_K(\tau) d\tau} 
		\end{eqnarray}
		and $c_k^- = c_k$ and $c_k^+ = c_k^\dagger$.
		Assuming that the contact between the molecule and the lead is established at $t=0$ and that the leads are initially in a thermal state corresponding to $H_K(0)$, 
		the correlation function can be expressed as
		\begin{eqnarray}
		      C_K^\pm(t, t')	&=&	
					      \frac{1}{2\pi} \int d\epsilon\  e^{\pm \frac{i}{\hbar}\epsilon (t-t')} \ \Gamma_K(t, t', \epsilon)\ f(\pm\epsilon, \pm\mu_K), \nonumber \\ \label{eq:correlation_function_evaluated}
		\end{eqnarray}
		where $f(\epsilon, \mu) = \left( 1 + \exp(\beta(\epsilon-\mu)\right)^{-1}$ is the Fermi distribution function with $\mu$ being the chemical potential, $\beta = \frac{1}{k_B T}$ with Boltzmann constant $k_B$ and temperature $T$.
		The construction of the hierarchical equations relies on a decomposition of the correlation function, 
		\begin{eqnarray}
			C_K^\pm(t, t') = \sum_p C_{Kp}^\pm(t, t'), 
		\end{eqnarray}
		which is closed with respect to taking the time derivative, that is 
		\begin{eqnarray}
			\partial_t C_K^\pm(t, t') = \sum_p \gamma_{Kp\pm}(t) C_{Kp}^\pm(t, t'),
		\end{eqnarray}
		where $\gamma_{Kp\pm}(t)$ is a set of parameters.
		This gives rise to the pole index $p_j\in \mathbb{N}$. 	%Mathematically, it is possible to work with an infinite number of poles, such that the decomposition of $C_K^\pm(t, t')$ does not impose any approximation. 
		In practical calculations, only a finite number of poles can be taken into account, such that the decomposition scheme has to be chosen carefully. Originally, the correlation function was decomposed into a set of exponential functions, using a Matsubara\cite{Mahan, Tanimura2006, Jin2008} or Pade\cite{Hu2010, Hu2011} decomposition scheme. Other approaches include the Chebyshev decomposition,\cite{Tian2012, Popescu2015, Popescu2016} or a general expansion in terms of a complete set of orthogonal functions.\cite{Tang2015}
% 		Recently, we have proposed an extended HQME method that can numerically treat an infinite number of poles.\cite{Erpenbeck2018_RSHQME}
		
		In the case of a time-dependent molecule-lead coupling, however, the resulting time-dependency of $\Gamma_K(t, t', \epsilon)$ directly enters the derivative $\partial_t C_K^\pm(t, t')$, complicating the construction of the hierarchy and effectively restricting time-dependent molecule-lead couplings to specific forms.
		Therefore, it is more convenient to construct the hierarchy from the function		
		\begin{eqnarray}
			\tilde C_K^\pm(t, t')	&=&	
						\int d\epsilon\  e^{\pm \frac{i}{\hbar}\epsilon (t-t')} \ V_{K}^*(t', \epsilon)\ f(\pm\epsilon, \pm\mu_K) .\nonumber \\
		\end{eqnarray}
		Expressing the Fermi function in terms of a Matsubara or Pade decomposition and performing the integral in the complex plane using Cauchy's theorem, $\tilde C_K^\pm(t, t')$ can be approximated by a sum over exponentials allowing for a systematic closure of the hierarchy.\cite{Yan2008, Zheng2012, Haertle2013a}
		Proceeding in this way, the objects contained in the HQME (\ref{eq:EQM_nth_tier}) assume the form
		\begin{subequations}
		\begin{eqnarray}
			\gamma_{a}(t)					&=& - \sigma \frac{i}{\hbar} \left(\epsilon_K(t) - \epsilon_K(0) + \mu_K + \frac{i \sigma \chi_p}{\beta} \right)  , \nonumber \\ \\ 
			\mathcal{C}_{a}(t) \rho^{(n)}(t)		&=& - \frac{2i\pi V_{K}(t)}{\beta}  \eta_p   \left\lbrace d^{\sigma}, \rho^{(n)}(t) \right\rbrace_{(-1)^{n+1}} , \nonumber \\  \\
			A_{K}^{\overline{\sigma}}(t) \rho^{(n)}(t) 	&=& V_{K}(t) \left\lbrace d^{\overline{\sigma}}, \rho^{(n)}(t) \right\rbrace_{(-1)^{n}} ,
		\end{eqnarray}
		\end{subequations}
		where $\overline{\sigma}= - \sigma$, $d^- = d$ and $d^+ = d^\dagger$. $\lbrace . , . \rbrace_-$ denotes the commutator, $\lbrace . , . \rbrace_+$ is the anti-commutator. These expressions are specific for the wide-band limit and the Pade decomposition \cite{Hu2010, Hu2011} used throughout this paper. How to calculate the Pade decomposition parameters $\eta_p$ and $\chi_p$ was for example demonstrated by \citet{Hu2011}
		Using the wide-band description of the leads within the HQME framework requires an additional auxiliary object in every tier, which is given by
		\begin{eqnarray}
			\rho_{a_1 \dots a_n (K, 0, \sigma)}^{(n+1)} 	&=&	- \frac{i\pi\hbar V_{K}(t)}{2} 
										\cdot \left\lbrace d^{\sigma}, \rho_{a_1 \dots a_n}^{(n)} \right\rbrace_{(-1)^{n+1}} . \nonumber\\
		\end{eqnarray}
% 		For technical details on the wide-band description within the HQME methodology, we refer to our recent publications Refs.\ \onlinecite{Erpenbeck2018, Erpenbeck2018_RSHQME} and the references therein.
		For technical details on the wide-band description within the HQME methodology, we refer to our recent publication Ref.\ \onlinecite{Erpenbeck2018} and the references therein.

		Several observables are of interest to study charge transport in molecular junctions. The most fundamental observable is the electrical current, which within the HQME framework, is given by
		\begin{eqnarray}
			I_{K}(t)	&=&	\frac{ie}{\hbar^2} \sum_{p \in {\text{poles}}} V_{K}(t) \text{Tr}\left( d \rho_{Kp+}^{(1)}(t) - d^\dagger \rho_{Kp-}^{(1)}(t) \right)  , \nonumber \\ \label{eq:current}
		\end{eqnarray}
		for lead $K$. Here Tr denotes the trace over the system degrees of freedom.
		When considering a time-dependent influence on the molecular junction, it is also insightful to study the charge pumped by the time-dependency, which is calculated as
 		\begin{eqnarray}
 			Q_{K} &=& \int dt \ \Big( I_{K}(t) - I_{K 0}(t) \Big). \label{eq:charge_pumped}
 		\end{eqnarray}
 		Thereby, $I_{K}(t)$ is the current in presence of the time-dependent influence, whereas $I_{K 0}(t)$ is the current without it. This expression is particularly useful when considering the influence of time-dependent pulses.

		In the HQME (\ref{eq:EQM_nth_tier}), the auxiliary density operators of different tiers are coupled. In general, this results in an infinite hierarchy of coupled differential equations, which has to be truncated in a suitable manner for applications.\cite{Tanimura1991, Yan2004, Xu2005, Schroeder2007} 
		As we are studying a noninteracting systems by means of the electronic current, which is a single particle observable, the hierarchy terminates after the $2$nd-tier.\cite{Yan2008, Karlstrom2013}
		Within the wide-band limit, it is sufficient to only include the $1$st-tier auxiliary density matrices and still obtain numerically exact results.\cite{Croy2009, Zheng2010, Kwok2014, Leitherer2017}

		At this point, we want to remark on the time-dependencies that can be treated within the HQME framework. Any time-dependency of the molecular Hamiltonian naturally enters the equations of motion (\ref{eq:EQM_nth_tier}) via the commutator with the molecular Hamiltonian $H_{\text{M}}(t)$. Time-dependent energies of the noninteracting leads can also be treated, if the time-dependency is the same for all lead states, and enter the equations of motion via $\gamma_{a}(t)$. 
		Formally neither the chemical potential nor the temperature can be time-dependent in the formulation of the HQME theory as these quantities do not enter any Hamiltonian operator. 
		However, the chemical potentials only enter the equations of motion (\ref{eq:EQM_nth_tier}) via $\gamma_{a}(t)$, and are thus always related to the energy of the electronic states in the leads, such that it is possible to express the influence of time-dependent chemical potentials by time-dependent lead energies.

\section{Results}\label{sec:results}
	As an illustrative example, we study in this section the pumping of charge by the modulation of the molecule-lead coupling strength. To this end, we consider a model where the coupling between the molecule and the right lead $V_\text{R}$ is constant in time, while the coupling between the molecule and the left lead is modulated by a sinusoidal Gaussian pulse of frequency $\omega$ and duration $25$fs,
	\begin{eqnarray}
		  V_\text{L}(t) &=& V_\text{L0} \cdot \left(1 + 0.9 \cdot \sin\big(\omega(t-90\text{fs})\big)  e^{-\left(\frac{t-90\text{fs}}{25\text{fs}}\right)^2}\right) . \nonumber \\ \label{eq:results_form_1}
	\end{eqnarray}
	This time-dependent coupling influences the transport processes between the molecule and the left lead, which also depend on the chemical potential of the left lead $\mu_{\text{L}}$. The transport processes between the molecule and the right lead are not directly influenced by the time-dependency of $V_\text{L}(t)$.
	Therefore, we set the chemical potential of the right lead to the constant value of $\mu_{\text{R}}=-0.3$ V throughout this section, such that the right lead serves as a sink for the electrons. The chemical potential of the left lead $\mu_{\text{L}}$ is variable enabling the regulation of transport processes between the left lead and the molecule. % that are energetically possible.
	For the molecule, we consider a single electronic level model  with energy $\epsilon_0 = 0.4$ eV. The leads are modeled in the wide-band limit, the temperature is set to $300$ K.
	It is noted that models where the left and the right molecule-lead couplings depend on time were already investigated in the realm of electron pumps and turnstile devices.\cite{Splettstoesser2005, Braun2008, Cavaliere2009, Romeo2010, Croy2012, Croy2012_b, Kwapinski2011}
	The model parameters used here are representative for molecular junctions.\cite{Elbing05, Benesch2008, Benesch2009, Ballmann2010, Arroyo2010, Ballmann2012, Erpenbeck2015, Erpenbeck2016, Schinabeck2016} 
	
\begin{figure*}
		\begin{minipage}{.45\textwidth}
			\vspace*{-0.4cm}
			\includegraphics[width=\textwidth]{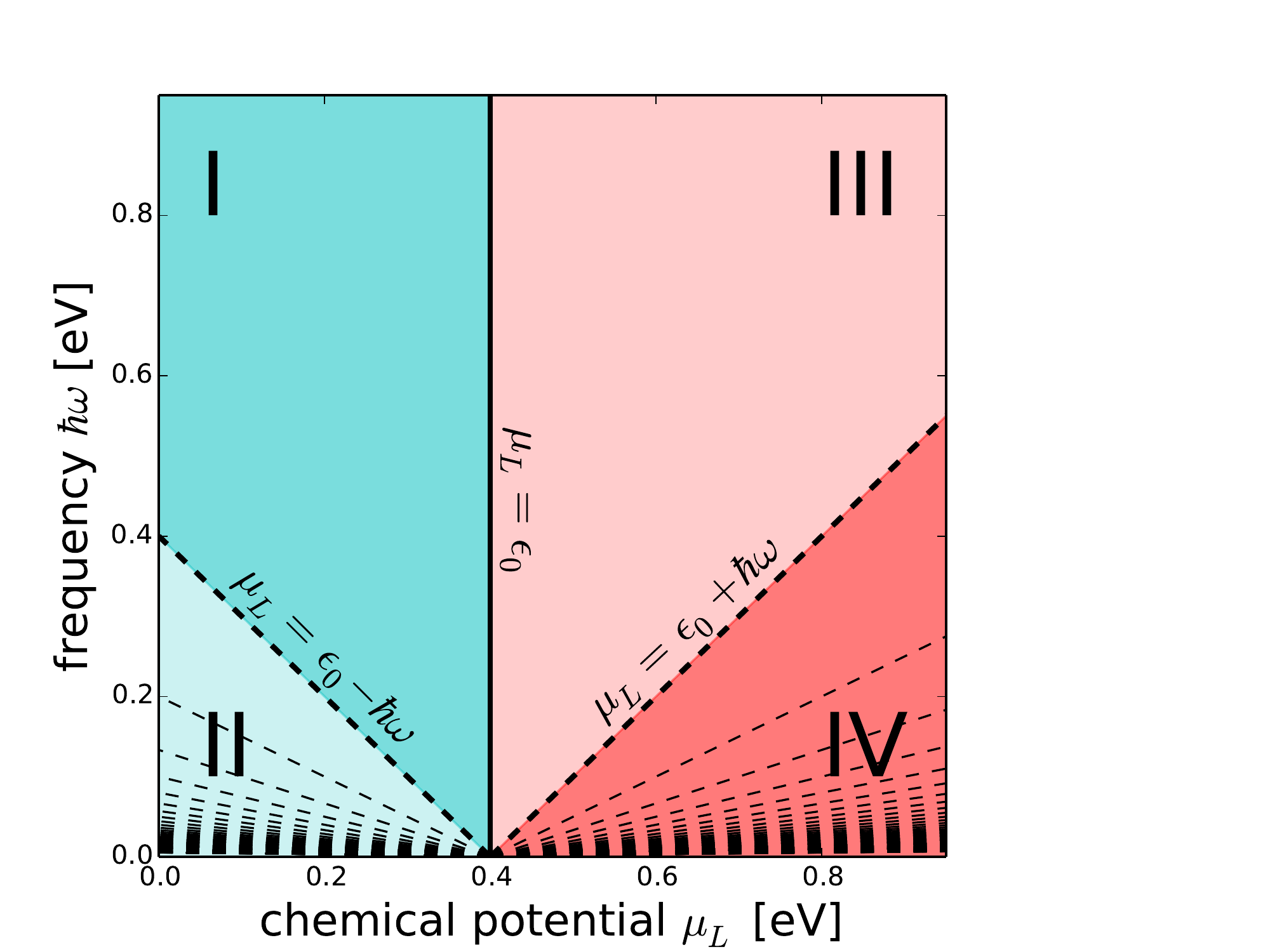}
			\caption{Classification of the parameter space in terms \\of transport processes that are energetically possible.}\label{fig:classification}
		\end{minipage}
		\begin{minipage}{.45\textwidth}
 			\vspace*{1.2cm}
			\includegraphics[width=\textwidth]{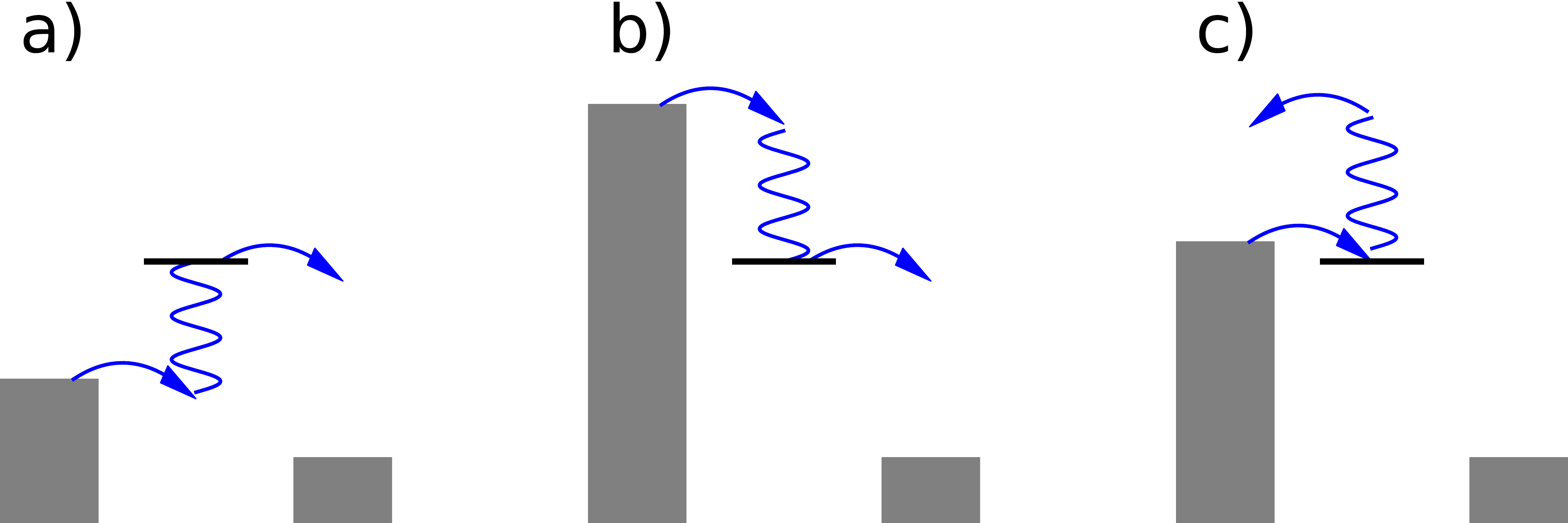}
			\vspace*{1.36cm}
			\caption{Transport processes including the emission or absorption of one virtual photon.}\label{fig:processes}
		\end{minipage}
		\begin{minipage}{.45\textwidth}
				\vspace*{1cm}
				a) symmetric coupling scenario, weak molecule-lead coupling $V_\text{L0} = V_\text{R} = 0.03$eV
				\includegraphics[width=\textwidth]{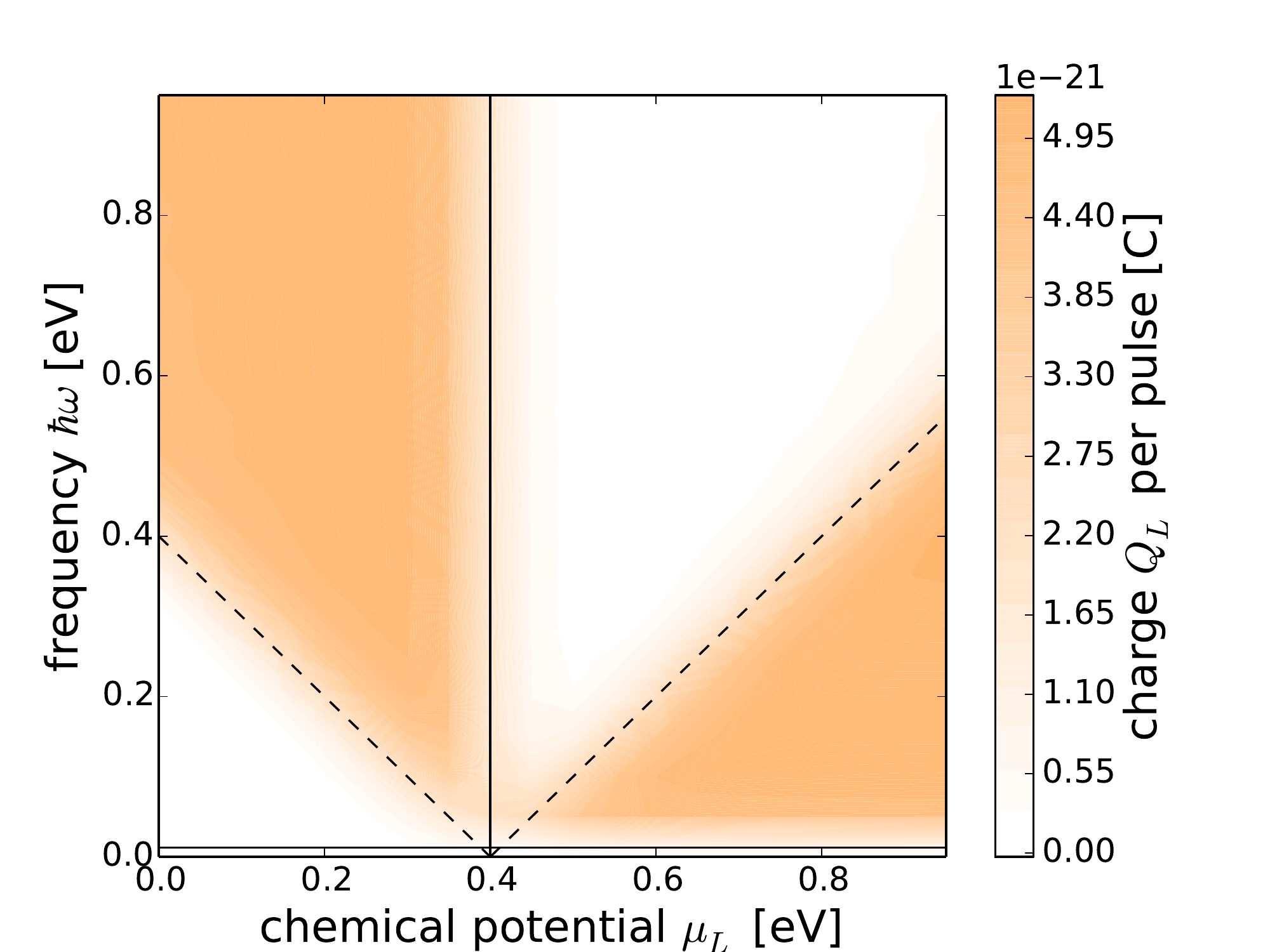}
		\end{minipage}
		\begin{minipage}{.45\textwidth}
				\vspace*{1cm}
				b) symmetric coupling scenario, strong molecule-lead coupling $V_\text{L0} = V_\text{R} = 0.1$eV
				\includegraphics[width=\textwidth]{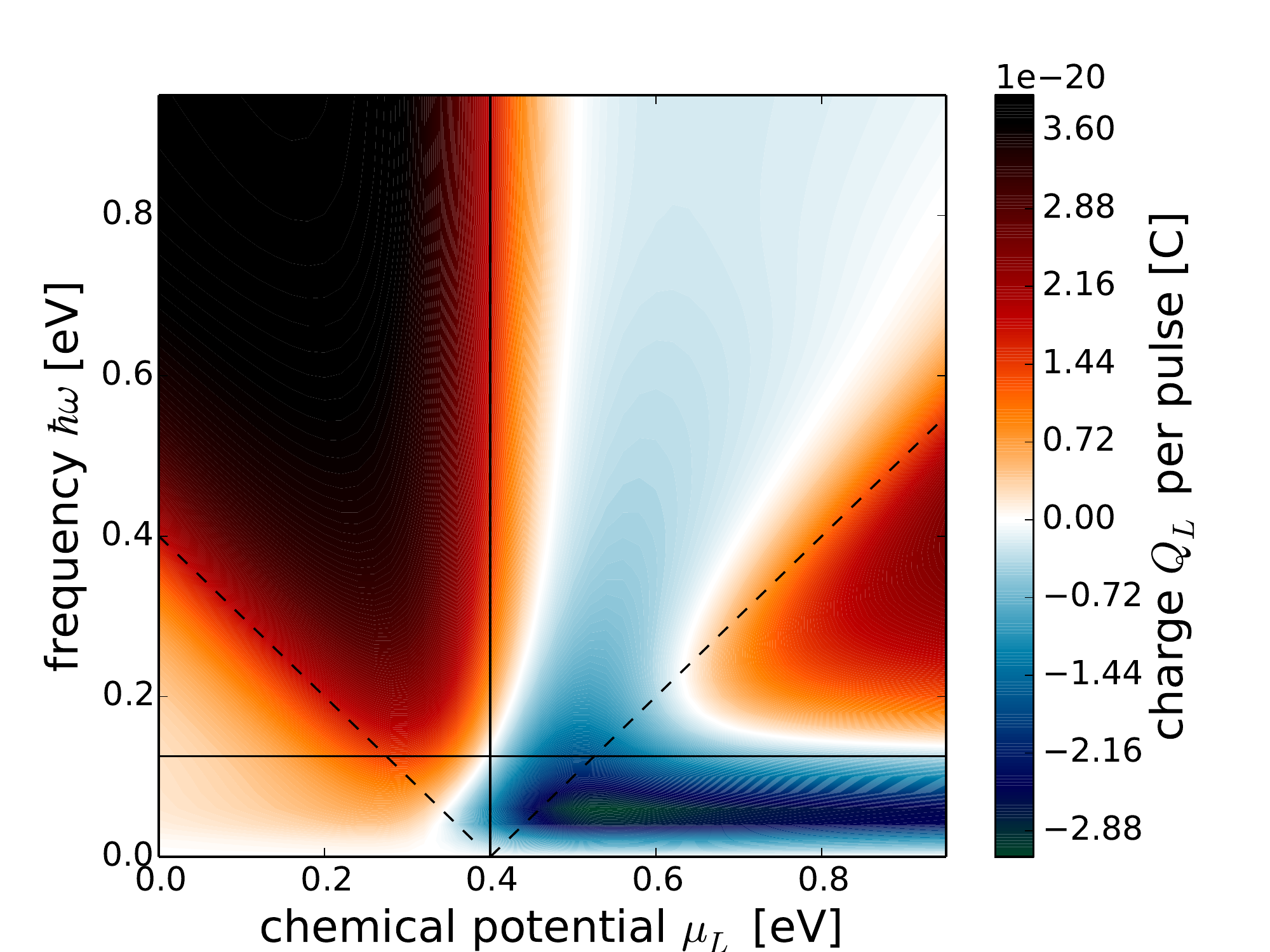}
		\end{minipage}	
		\begin{minipage}{.45\textwidth}
				\vspace*{0.7cm}
				c) asymmetric coupling scenario, $V_\text{L0} = 0.1$eV, $V_\text{R} = 0.03$eV
				\includegraphics[width=\textwidth]{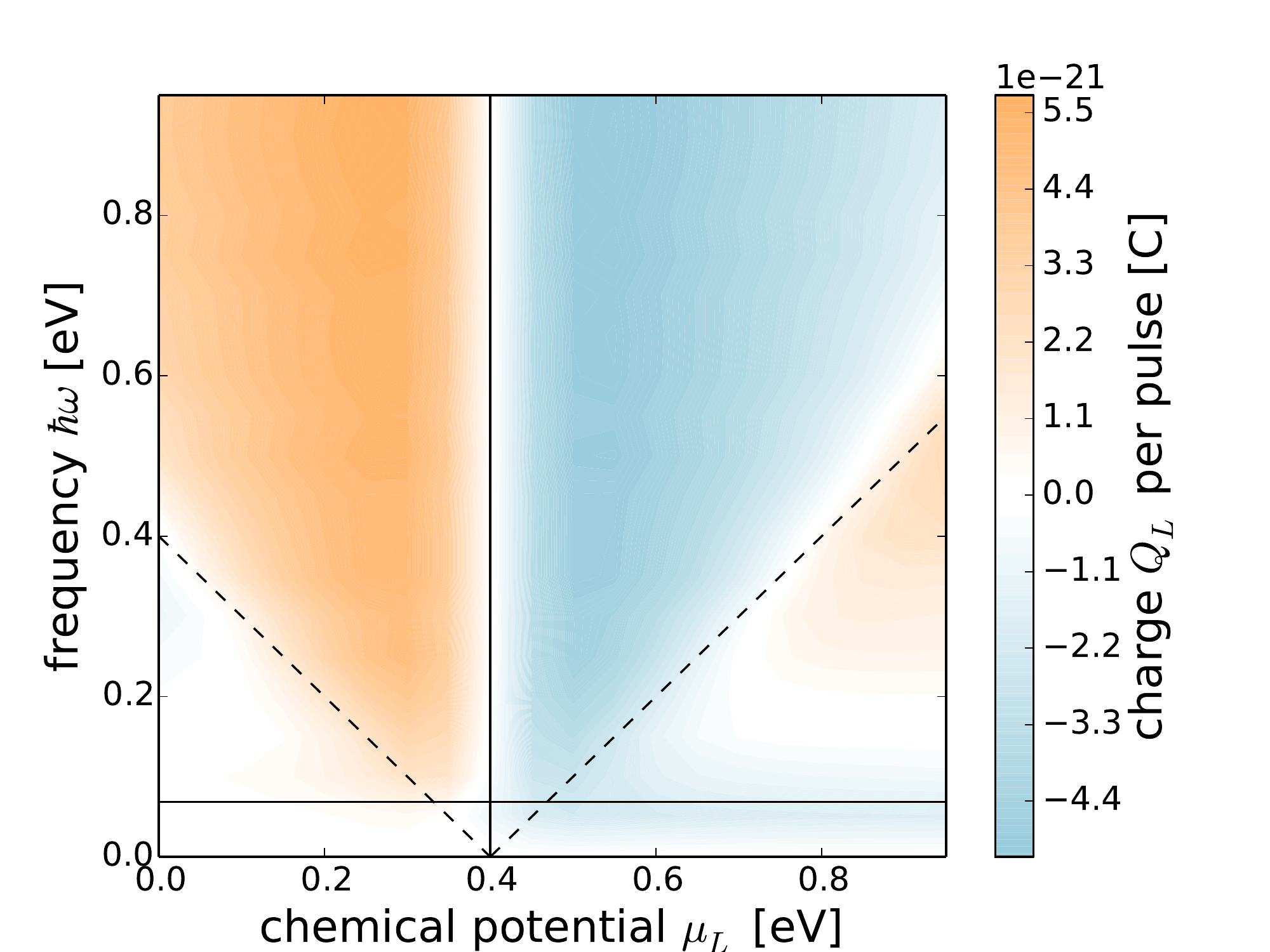}
		\end{minipage}
		\begin{minipage}{.45\textwidth}
				\vspace*{0.7cm}
				d) asymmetric coupling scenario, $V_\text{L0} = 0.03$eV, $V_\text{R} = 0.1$eV
				\includegraphics[width=\textwidth]{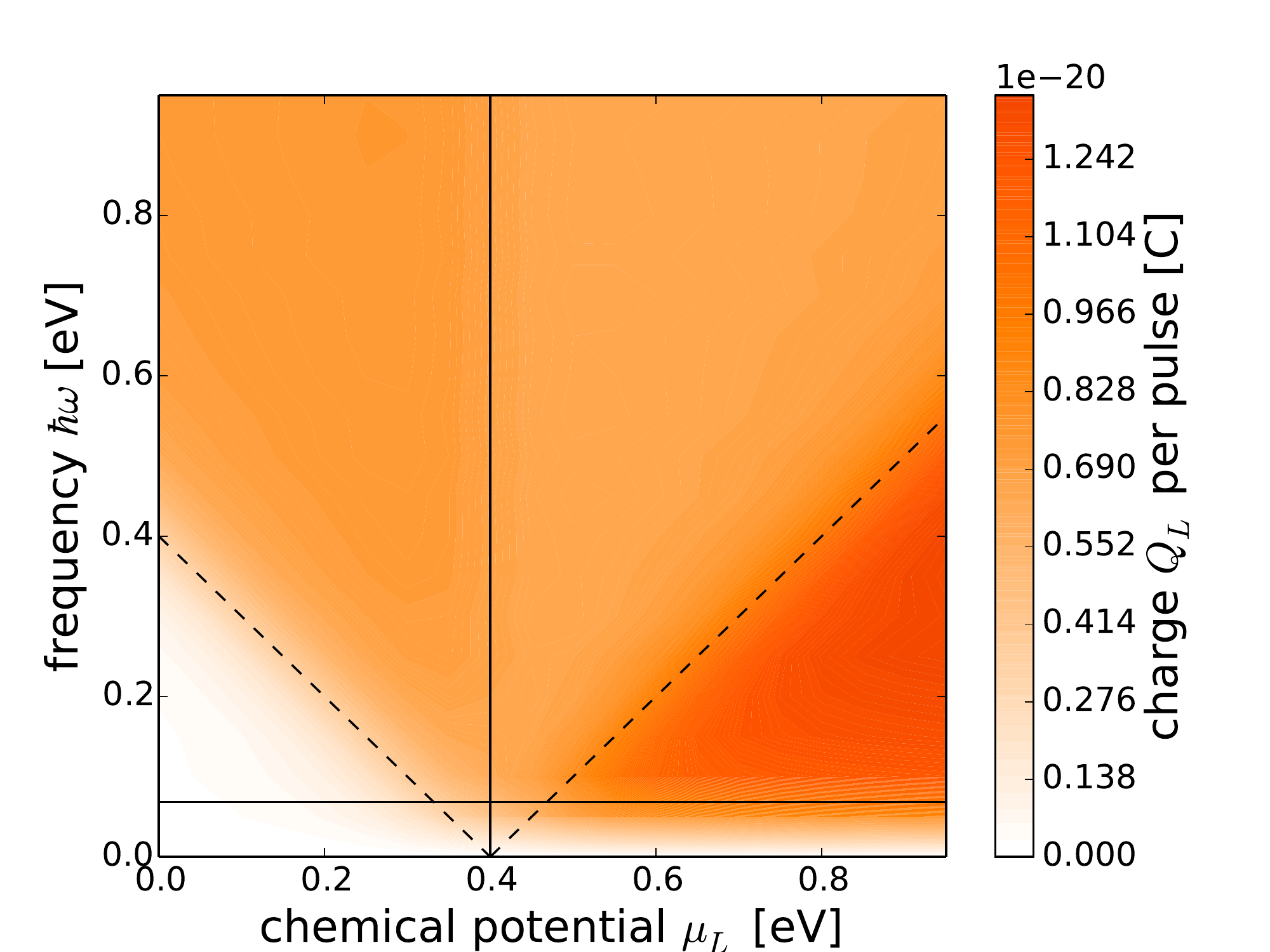}
		\end{minipage}
		\caption{Change in the electronic current according to Eq.\ (\ref{eq:charge_pumped}) upon a time-dependent molecule-lead coupling of the form given in Eq.\ (\ref{eq:results_form_1}). Vertical black lines indicate the transition between non-resonant and resonant transport regime, horizontal black lines indicate the adiabatic regime, inclined dashed lines mark the one-photon resonances.}
		\label{fig:1}
	\end{figure*}

	In the following, we will discuss the charge pumped by the time-dependent molecule-lead coupling, given by Eq.\ (\ref{eq:results_form_1}), as a function of chemical potential $\mu_\text{L} \in [0\text{ eV}, 1\text{ eV}]$ and pumping energy $\hbar\omega \in [0\text{ eV}, 1\text{ eV}]$ for different molecule-lead coupling strengths and scenarios. Before studying the actual data, we start by giving an overview of the different transport processes that are possible depending on $\mu_{\text{L}}$ and $\hbar\omega$. Fig.\ \ref{fig:classification} provides a classification of the parameter space into four regions in terms of transport processes that are energetically possible. This map is used subsequently for the interpretation of the numerical results.
	The blue areas I and II define the non-resonant transport regime, whereas the red regions III and IV correspond to the resonant transport regime.
	The resonance condition $\mu_{\text{L}} = \epsilon_0-\hbar\omega$ separates the regions I and II from each other. In region II it is energetically not possible to populate the molecular electronic level upon the absorption of one virtual photon, as depicted in Fig.\ \ref{fig:processes}a. The resonance conditions for multi-photon processes are also located in region II and are highlighted by dashed lines. Likewise, regions III and IV are separated by the resonance condition $\mu_{\text{L}} = \epsilon_0+\hbar\omega$ such that only in region IV the population of the molecular electronic level upon the emission of one virtual photon, as depicted in Fig.\ \ref{fig:processes}b, is possible. Moreover, there is also the possibility for electron-hole pair creation processes upon the absorption of one virtual photon. The corresponding process is depicted in Fig.\ \ref{fig:processes}c. Although this process is energetically possible in the regions I, II and III, it crucially depends on the population of the molecule, such that the process is most influential in region III.

	Fig.\ \ref{fig:1} depicts numerical data obtained for the model system, which provide an overview of the charge pumped by the time-dependent molecule-lead coupling strength, Eq.\ (\ref{eq:results_form_1}), as a function of chemical potential $\mu_\text{L}$ and pumping frequency $\omega$ for different molecule-lead coupling scenarios.
	Figs.\ \ref{fig:1}a and b correspond to the case of symmetric molecule-lead coupling, that is $V_\text{L0} = V_\text{R}$, for weak and strong molecule-lead coupling, respectively. Figs.\ \ref{fig:1}c and d depict the asymmetric scenario, where the coupling to either the left or the right lead is dominant.
	
	Fig.\ \ref{fig:1} reveals that the amount and the direction of the charge pumped depends crucially on the molecule-lead coupling strength, the specific scenario as well as the transport regime. The dashed vertical line indicates the chemical potential $\mu_{\text{L}}=\epsilon_0$, which separates the resonant from the non-resonant transport regime. Further, we find that the resonance conditions $\mu_{\text{L}}=\epsilon_0\pm\hbar\omega$, which are highlighted by the inclined dashed lines, are fundamental for the charge pumped. Albeit the HQME method includes any order in the molecule-lead coupling strength, we do not observe indications of higher order photon processes, even for the strong molecule-lead coupling case. 
	As the sine-modulation in $V_\text{L}(t)$ is de facto composed of two exponentials, this finding can be explained by the reasoning of Sec.\ \ref{sec:time-dep}, assuming the interplay between the two exponentials is negligible.
	Finally,  also the time-scale of the tunneling electrons in comparison to the pumping frequency is relevant. The horizontal black lines mark the energy condition $\hbar\omega = \Gamma_\text{L}(0)+\Gamma_\text{R}$, which we use as rough separation between the adiabatic ($\hbar\omega < \Gamma_\text{L}(0)+\Gamma_\text{R}$) and the anti-adiabatic regime ($\hbar\omega > \Gamma_\text{L}(0)+\Gamma_\text{R}$). We will discuss these observations and their associated transport mechanisms in the following.
	
	First, we consider the results of the weakly coupled, symmetric system in Fig.\ \ref{fig:1}a. In this case, there is no pronounced adiabatic regime due to the weak coupling. The time-dependent molecule-lead coupling strength either leads to an increased current, which corresponds to a positive charge pumped, or has no actual influence on the current. 
	The behavior of the charge pumped is well classified by the regions I -- IV introduced in Fig.\ \ref{fig:classification}.
	In regions I and IV, the time-dependency of $V_\text{L}(t)$ increases the current as a consequence of the additional transport processes that enhance the population of the molecular electronic level from the left (Figs.\ \ref{fig:processes}a and b).
	In region III, the population induced by the absorption of a virtual photon (Fig.\ \ref{fig:processes}a) and the electron-hole pair creation process (Fig.\ \ref{fig:processes}c) are equally probable due to the symmetric molecule-lead coupling, such that there is no net current. In region II, there are no transport processes possible that include the time-dependent molecule-lead coupling.

	The strongly coupled, symmetric system is considered in Fig.\ \ref{fig:1}b.
	The charge pumped by a time-dependent molecule-lead coupling Eq.\ (\ref{eq:results_form_1}) is distinctively different from the weak coupling case in Fig.\ \ref{fig:1}a, thus establishing the necessity for a treatment of the molecule-lead coupling beyond lowest order as provided by the HQME approach. Generally, the resonance conditions used to classify the regimes I -- IV in Fig.\ \ref{fig:classification} are substantially broadened by the strong molecule-lead coupling. Also, there is a distinguished adiabatic transport regime displaying a different physical behavior for low pumping energies $\hbar\omega$.
	In contrast to the weak coupling case, the time-dependent molecule-lead coupling can result in an increase or a decrease in current. In the resonant transport regime, it is possible to switch between enhancing and diminishing the current upon a change in modulation frequency $\omega$. 
	
	Altogether, there are two regions in the resonant transport regime displaying negative pumped charges which we will explain in the following. For small, adiabatic molecule-lead driving frequencies $\omega$, the electronic response is fast compared to the time scale at which the molecule-lead coupling changes. Therefore, the electronic current assumes quasi-instantly the steady-state value for any given coupling strength. The steady state current depends in a nonlinear way on $V_\text{L}(t)$, such that the current integrated over the pulse duration is smaller compared to a system without time-dependent molecule-lead coupling.
	Furthermore, there is also a negative charge pumped for intermediate to high pumping frequencies $\omega$ in the resonant transport regime, which coincides roughly with region III.
	In this region, a negative charge is not explicable within the simple transport-process framework used so far. For the high molecule-lead coupling strength considered here, the interplay between the constant and the time-dependent parts of $V_\text{L}(t)$ as specified in Eq.\ (\ref{eq:results_form_1}), along with an extended amount of states in the leads contributing to transport, give rise to a nontrivial characteristics of the charge pumped. For the specific system parameters considered here, this results in a negative current in region III, but also the charge pumped in regions I and IV is smaller in magnitude than expected from rate theory.

	Finally, we consider the asymmetric coupling scenarios in Figs.\ \ref{fig:1}c and d. In these cases, the classification introduced in Fig.\ \ref{fig:classification} provides a reasonable ground for interpreting the data. The broadening due to the molecule-lead coupling plays a minor role and so does the adiabatic regime. Overall, the system more strongly coupled to the left lead in Fig.\ \ref{fig:1}c behaves similar to the symmetric model in Fig.\ \ref{fig:1}a. However, for the coupling scenario in Fig.\ \ref{fig:1}c, the time-dependent molecule-lead coupling results in a negative contribution to the current localized in region III. Due to the asymmetric molecule-lead coupling scenario, the bottleneck for the transport is the interface between the molecule and the right lead. This results in a highly populated molecule in the resonant transport regime, breaking the symmetry between the population upon the absorption of a virtual photon (Fig.\ \ref{fig:processes}a) and the depopulation of the molecule via the electron-hole pair creation process (Fig.\ \ref{fig:processes}c), which establishes as a negative current.
	For the system more strongly coupled to the right lead (Fig.\ \ref{fig:1}d), however, we always find an enhancement in current due to the time-dependent molecule-lead coupling. In this coupling scenario, the bottleneck for the transport is the interface between the molecule and the left lead, leading to an almost empty molecular electronic state in the resonant transport regime. As a consequence, processes populating the molecule upon interaction with one virtual photon are enhanced, resulting in a positive current.

\section{Conclusion}\label{sec:conclusion}
	We have investigated electron transport in models for molecular junctions in the presence of time-dependent external influences. 
	Complementing previous work on driven transport, we have shown that different time-dependent model parameters encode different physical processes. We have, furthermore, demonstrated that the HQME method is capable of treating a variety of different time-dependent model parameters on a numerically exact level and explicitly considered a representative model system with a time-dependent molecule-lead coupling for a wide range of molecule-lead coupling strengths and scenarios. We found that the system displays characteristic transport processes associated with time-dependent molecule-lead couplings, which depend in a unique way on the coupling strength and which cannot be described completely using traditional lowest order approaches.

\section*{Acknowledgement}
	This work was supported by the German Research Foundation (DFG) through SFB 953 and a research grant.

\bibliography{Bib}

\end{document}